\documentclass[12pt]{article}

\usepackage{amsmath,amssymb}

\usepackage{graphicx}

\global\arraycolsep=1pt
\numberwithin{equation}{section}
\newcommand{\bel}[1]{\begin{equation}\label{#1}}
\newcommand{\bal}[1]{\begin{eqnarray}\label{#1}}
\newcommand{\be}{\begin{equation}}
\newcommand{\ee}{\end{equation}}

\newcommand{\dis}{\displaystyle}
\newcommand{\scr}{\scriptstyle}
\newcommand{\qq}{\qquad}

\renewcommand{\thefootnote}{\fnsymbol{footnote}}
\def\fnote#1#2{\begingroup\def\thefootnote{#1}\footnote{#2}\addtocounter{footnote}{-1}\endgroup}

\begin{document}
\begin{flushright}
OCU-PHYS 207 \\
hep-th/0610092
\end{flushright}
\vspace{20mm}

\begin{center}
{\bf\Large
New Example of Infinite Family of\\
Quiver Gauge Theories
}
\end{center}

\begin{center}
\vspace{15mm}
Takeshi Oota \fnote{$\dagger$}{
\texttt{toota@sci.osaka-cu.ac.jp}
}, and
Yukinori Yasui\fnote{$\ast$}{
\texttt{yasui@sci.osaka-cu.ac.jp}
}
\vspace{10mm}

\textit{
${{}^\dagger}$
Osaka City University
Advanced Mathematical Institute (OCAMI)\\
3-3-138 Sugimoto,Sumiyoshi,
Osaka 558-8585, JAPAN
}
\vspace{2mm}

${}^\ast$
\textit{
Department of Mathematics and Physics, Graduate School of Science,\\
Osaka City University\\
3-3-138 Sugimoto,Sumiyoshi,
Osaka 558-8585, JAPAN
}

\vspace{5mm}

\end{center}
\vspace{8mm}

\begin{abstract}
 We construct a new infinite family of quiver gauge theories
which blow down to the $X^{p,q}$ quiver gauge theories
found by Hanany, Kazakopoulos and Wecht. This family
includes a quiver gauge
theory for the third del Pezzo surface. We show,
using Z-minimaization, that these theories generically have
irrational R-charges.
The AdS/CFT correspondence
implies that
the dual geometries are irregular toric
Sasaki-Einstein manifolds, 
although we do not know the explicit metrics.
\end{abstract}

\vspace{25mm}

\newpage

\section{Introduction}

D3-branes at the tip of Calabi-Yau cones have been extensively
studied. The corresponding IIB supergravity solution in the near horizon
limit takes the form $AdS_5 \times X_5$,
where $X_5$ is a five dimensional Sasaki-Einstein
manifold. In dimension five, simply-connected regular Sasaki-Einstein
manifolds are classified \cite{BFGK}.
Indeed we have $S^5$, $T^{1,1}$ and the total space $S_k$ of the 
circle 
bundles $S_k \rightarrow dP_k~ (3 \le k \le 8)$ where $dP_k$ is a 
del Pezzo surface with a K\"ahler Einstein metric. It is known that
$S_k$ is diffeomorphic to the k-fold connected sum 
$\sharp k(S^2 \times S^3)$.
Recently, an infinite family of irregular toric 
Sasaki-Einstein manifolds $Y^{p,q}$ with topology
$S^2 \times S^3$ was constructed \cite{GMSW,MS}. 
Especially, $Y^{2,1}$ is the horizon of the complex
cone over the first del Pezzo surface $dP_1$. Also, the existence
of irregular
toric Sasaki-Einstein manifolds $X^{p,q} \simeq (S^2 \times S^3) \sharp
(S^2 \times S^3)$ which blow down to
$Y^{p,q}$ and $Y^{p,q-1}$ by Higgsing was conjectured in \cite{HKW}.
These are considered to be extension of the
Sasaki-Einstein manifold $X^{2,1}$
over the second del Pezzo surface
$dP_2$\cite{FOW}. For other new 
Sasaki-Einstein manifolds see \cite{CLPP1,CLPP2,CLP} and also \cite{BG}.

The AdS/CFT correspondence states that IIB string theory on
$AdS_5 \times X_5$ is dual to a four-dimensional $\mathcal{N}=1$
quiver gauge theory. Given a toric Sasaki-Einstein manifold, one can
determine the corresponding quiver gauge theory  by using the
brane tiling (dimer) construction\cite{HK,FHKVW,FHMSVW,HV,FHKV,HHV,FV,I1,I2}. 
In Figures 1, 2 and 3 , we present some data of quiver gauge
theories for the del
Pezzo surfaces $dP_k~(k=1,2,3)$, corresponding to 
$Y^{2,1}, X^{2,1}$  and $S_1$
\cite{FHH,FHSU,FHU,BGLP,BBC,FHH2,BP,FHHU,FFHH,FFHH2,FH,IW}.  
These quiver theories have been extended to 
the general $Y^{p,q}$\cite{BFHMS,BFZ} 
and $X^{p,q}$ theories\cite{HKW,BZ1,BZ2}.

In this paper we construct an infinite family of $\mathcal{N}=1$ 
quiver gauge theories which blow down to $X^{p,q}$.
This construction generalizes the $dP_3$ quiver gauge theory,
corresponding to $S_1 \equiv Z^{2,1}$.
The dual geometries are  irregular toric Sasaki-Einstein manifolds
which are diffeomorphic to $\sharp 3(S^2 \times S^3)$.
We denote as $Z^{p,q}$ assuming 
they exist. It should be mentioned that the existence of $Z^{p,q}$
was suggested in a recent paper\cite{ABCC}.

In section 2 we describe the brane tiling construction of
the $Z^{p,q}$ theories. The explicit examples are given in
Figures 5-14. In section 3 we determine the $R$-charges for 
these theories by using Z-minimaization.

\section{Un-Higgsing $X^{p,q}$}

The brane tiling for $X^{p,q}$ was given in \cite{BZ1}.
We use the convention of \cite{BZ1} for the brane tiling.
Some data for the $X^{p,q}$ quiver gauge theory are summarized in
Figure 4. 
We take $n=2q-1$, $m=p-q$, $j=1,2,\dotsc, q-1$ and 
$k=n+1,n+2,\dotsc, n+m-1$. Then, the brane tiling contains 
$2q-1$ hexagons and $p-q+1$ cut hexagons.
The $R_{i}+R_{i+1}+ \cdots$ on  edges represent
$R$-charges: at every vertex the sum of $R$-charges is 2 and
for every face the sum of $R$-charges  is equal to the number
of edges minus 2.

We can un-Higgs $X^{p,q}$ to $Z^{p,q}$ by
cutting the $i$-th hexagon horizontally $(1 \leq i \leq n$).
This procedure introduces one new edge with weight $-w^{-1}$
which connects the $(i+2)$-th white node
and the $i$-th black node.
If we replace $0$ at $(i+2,i)$ element of the Kasteleyn matrix
for $X^{p,q}$ with $-w^{-1}$, we  obtain  the Kasteleyn matrix
for $Z^{p,q}$.
The determinant is modified to contain the term $\pm w^{-1} z$,
which corresponds to an additional node (see next section for details):
\be
V_6 = (1, w_6), \qq w_6 = (2,1).
\ee
The perfect matchings which correspond to the node $(2,1)$
are given by
\be
P^{(i)}_6 = \left(
\begin{array}{c|c@{\hspace{1em}}c@{\hspace{0.5em}}cc|c|cc@{\hspace{1em}}c@{\hspace{0.5em}}cc}
 & {\scr 1} & {\scr \dotsm} & {\scr \dotsm} & {\scr i} 
& {\scr i+1} &
{\scr i+2} & {\scr i+3} & {\scr \dotsm }& {\scr \dotsm }& 
{\scr n+m+1} \\ \hline
{\scr 1} & 0 & & & & & & & & & z \\
 & 1 & 0 & & & & & & & & \\ & & \ddots & \ddots & & & & & & & \\
{\scr i} & & & 1 & 0 & & & & & & \\ \hline
{\scr i+1} & & & & & 1 & 0 & & & & \\
{\scr i+2} & & & & -w^{-1} & 0 & 0 & & & & \\ \hline
{\scr i+3} & & & & & & 1 & 0 & & & \\
{\scr \vdots} & & & & & & & 1 & 0 & & \\
{\scr \vdots} & & & & & & & & \ddots & \ddots & \\
{\scr n+m+1} & & & & & & & & & 1 & 0 \\
\end{array}
\right).
\ee
The matrices $P^{(i)}_6~ (i=1,2,\cdots, n)$ have non-zero elements
$K_{ab}$ at the edges
$\stackrel{a}{\circ}\>\!\!\!-\!\!\!-\!\!\!\stackrel{b}{\bullet}$
connected by bold lines (see Figures 7, 10, 12 and 14), and the values $K_{ab}$ are
decided according to a rule of the Kasteleyn matrix. 
The charge $R_6$ is added to the corresponding edges.

The contents of the minimal toric phase are summarized as follows:
\begin{center}
\begin{tabular}[ht]{|c|c|c|c|} \hline
 & $N_g$ & $N_f$ & $N_W$ \cr \hline \hline
$Y^{p,q}$ & $2p$   & $4p+2q$   & $2p+2q$ \\
$X^{p,q}$ & $2p+1$ & $4p+2q+1$ & $2p+2q$ \\
$Z^{p,q}$ & $2p+2$ & $4p+2q+2$ & $2p+2q$ \\ \hline
\end{tabular}
\end{center}

Here $N_g$ represents the number of gauge groups, $N_f$ the number of bifundamental fields, $N_W$
the number of the interaction terms in the superpotential.
These correspond to the numbers of faces, edges and nodes on the
brane tiling, respectively.

In the following we
describe the brane tiling construction of the $Z^{3,1}$ and $Z^{3,2}$
theories explicitly.

\subsection{$Z^{3,1}$}

The brane tiling for $X^{3,1}$ is given by Figure 5.
In this case, there is only one hexagon that we can put a cut in.
We have drawn the resulting brane tiling for $Z^{3,1}$ in Figure 6
together with the corresponding quiver diagram and toric diagram.
The perfect matching in Figure 7 is given by the matrix
\be
P^{(1)}_6(Z^{3,1})
= \left(
\begin{array}{c@{\hspace{0.5em}}|c@{\hspace{1em}}c@{\hspace{1em}}c@{\hspace{1em}}c}
 & {\scr 1} & {\scr 2} & {\scr 3} & {\scr 4} \\ \hline
{\scr 1} & 0 & 0 & 0 & z \\
{\scr 2} & 0 & 1 & 0 & 0 \\
{\scr 3} & - w^{-1} & 0 & 0 & 0 \\
{\scr 4} & 0 & 0 & 1 & 0 \\
\end{array}
\right).
\ee

\subsection{$Z^{3,2}$}
The brane tiling for $X^{3,2}$ is given by Figure 8.
In this case we have three types of cuttings corresponding to
three hexagons. These lead to different brane 
tilings and quiver diagrams for $Z^{3,2}$, although
their toric diagrams are equivalent, as shown in Figures 8-14.

The superpotentials contain the terms with the following degree:

\begin{tabular}{|c|c|c|c|c|} \hline
         & cubic & quartic & quintic & sextic \cr \hline \hline
case $1$ & $5$ & $4$ & $1$ & $0$ \\
case $2$ & $6$ & $2$ & $2$ & $0$ \\ 
case $3$ & $6$ & $3$ & $0$ & $1$ \\ \hline
\end{tabular}

These provide an example of toric duality and
 are connected by Seiberg duality \cite{FHH,FHH2,BP,FFHH}.

\section{ Z-minimization }

The toric diagram for $Z^{p,q}$ can be obtained
by adding one vertex to the toric diagram for $X^{p,q}$,
as shown in Figures 15 and 16.
The six vectors $ V_i=(1, w_i)$ for $Z^{p,q}$ are written as
\begin{eqnarray}
w_1&=&(1,p),~~w_2=(0,p-q+1),~~w_3=(0,p-q),\nonumber \\
w_4&=&(1,0),~~w_5=(2,0),~~w_6=(2,1),
\end{eqnarray}
where $p$ and $q$ are integers with $0< q < p$. 
Let us determine the Reeb vector $b=(3,x,y)$ for $Z^{p,q}$
using the method of \cite{MSY}, Z-minimization. 
Then, we obtain $x=3$ and the remaining component $y$ is given by
a root of the polynomial 
\begin{equation}
P(y)=2 y^3-9 p q y^2-9p^2(2 p-3 q)y+27 p^3(p-q).
\end{equation}
This polynomial has three real roots $y=y_i~(i=1,2,3)$.
By $P(0)=27p^3(p-q)>0 $ and $P(3 p)=-27p^3(p+q-2)<0$
they satisfy $y_1<0<y_2<3 p<y_3$. 
We find that the correct root is a middle one $y_2$, which is explicitly given by
\begin{equation}\label{y2}
y_2=\frac{3 p q}{2}-3 p \sqrt{q^2-2 q+(4/3)p}
\cos \frac{\theta+\pi}{3},
\end{equation}
where the angle $\theta ~(0 \le \theta \le \pi/2)$ is calculated as
\begin{equation}
\tan \theta = \frac{\sqrt{D}}{(q-1)(q^2-2 q+2 p)}
\end{equation}
with
\begin{equation}
D=\frac{64}{27}p^3+\frac{4}{3}p^2(q+1)(q-3)-4 p q(q-2)-q^2(q-2)^2.
\end{equation}
The R-charges can be computed from volumes of certain calibrated
submanifolds $\Sigma_i$ in $Z^{p,q}$:
\begin{equation}
R_i=\frac{\pi vol(\Sigma_i)}{3 vol(Z^{p,q})},
\end{equation}
where
\begin{equation}
vol(\Sigma_i)=\frac{2(V_{i-1},V_i,V_{i+1})\pi^2}
{(b,V_{i-1},V_i)(b,V_{i},V_{i+1})},~~
vol(Z^{p,q})=\frac{\pi}{6}\sum_{i=1}^{6}vol(\Sigma_i).
\end{equation}
Explicitly we have the following volumes
\begin{eqnarray}
vol(\Sigma_1)&=&\frac{2(p+q-2)\pi^2}{(3 p-y_2)^2},~~
vol(\Sigma_2)=vol(\Sigma_6)=\frac{2 \pi^2}{3(3 p-y_2)},\nonumber \\
vol(\Sigma_3) &=& vol(\Sigma_5)=\frac{2 \pi^2}{3 y_2},~~
vol(\Sigma_4)=\frac{2(p-q) \pi^2}{y_2^2}
\end{eqnarray}
and
\begin{equation}
vol(Z^{p,q})=
\frac{9 p^3-9p^2 q+6 p q y_2-2 y_2^2}{3 y_2^2(3 p-y_2)^2}\pi^3.
\end{equation}
It should be noticed that\footnote{We calculate as
$vol(Z^{p,1})/\pi^3=8(2p-1)/(27 p^2)<1$, and the successive inequalities
can be evaluated by perturbation.}
\begin{equation}
1 > \frac{vol(Z^{p,1})}{\pi^3} > \frac{vol(Z^{p,2})}{\pi^3} > \cdots.
\end{equation}
This implies the inequality of the central charge 
\begin{equation}
a=\frac{\pi^3 N^2}{4 vol(Z^{p,q})} > \frac{N^2}{4},
\end{equation}
where $N$ is the number of D3-branes and $N^2/4$ the central charge
of $\mathcal{N}=4$ Yang-Mills theory.

The root $y_2$ \eqref{y2} is generically an irrational number and hence
$Z^{p,q}$ are irregular Sasaki-Einstein manifolds. 
As a special case we have a rational number 
$y_2=3 p/2$ for $q=1$.
 For
the toric diagram 
with six external lines, the third homology 
is given by $H_3(Z^{p,q})={\bf{Z}}^3$\cite{HKW}. 
It turns out that from the work of Smale\cite{S} 
$Z^{p,q}$ must be diffeomorphic to $\sharp 3(S^2 \times S^3)$
\footnote{In \cite{BGN} it has been shown that
$\sharp 3(S^2 \times S^3)$ admits an
infinite family of non-regular Sasaki-Einstein structures.
We do not know whether $Z^{p,q}$ are equivalent to those.}.

The cone of a Sasaki-Einstein manifold is Ricci-flat K\"ahler,~i.e,
Calabi-Yau. As described in \cite{MSY}, the cone metric
in the symplectic coordinates is given by a symplectic potential.
We found that for $Y^{p,q}$ the symplectic potential
takes
very simple form \cite{OY}. At present the Sasaki-Einstein metrics
on $X^{p,q}$ and $Z^{p,q}$ are not constructed or even proved to
exist for generic integers $p$ and $q$. It is expected
that the symplectic approach can be useful to study these metrics.

Before concluding this paper, we comment on the
the generating function $f$ for the single-trace
gauge invariant operators \cite{BFHH,MSY2}.
In the notation of \cite{BFHH}, the toric diagram for $Z^{p,q}$
is obtained by adding the point $(1,1,1)$
to the one for $X^{p,q}$ .
One more triangle with vertices $(1,0,1)$, $(1,1,1)$ and $(0,p,1)$
gives a new term to the generating function 
(see eq.(3.22) and the next equation of \cite{BFHH}),
\be
f(x,y,z; Z^{p,q}) = f(x,y,z; X^{p,q})
+ \frac{1}{\dis \left( 1 - \frac{x^p y}{z^p} \right)
\left( 1 - \frac{z}{x} \right) \left( 1 - \frac{z^p}{x^{p-1} y} \right)}.
\ee
Taking the limit \cite{MSY2}
\be
V(b)= \lim_{t \to 0} t^3 f(e^{-b_1t},e^{-b_2 t},e^{-b_3 t} ; Z^{p,q})
\ee
we have
\begin{eqnarray}
V(b)&=&\frac{-(p-1)b_1+p b_3}
{b_1 b_2 (b_1-b_3)((p-1)b_1+b_2-p b_3)}\nonumber \\
    &-&\frac{(p-1)b_1+p b_3}
{b_1 (b_1+b_3)((p-q)b_1+b_2)((q-1)b_1-b_2+p b_3)},
\end{eqnarray}
which reproduces the volume $vol(Z^{p,q})/\pi^3$ 
for the Reeb vector $(b_1, b_2, b_3)=(0, y_2, 3)$.


{\bf{Acknowledgements}}

We would like to thank K. Maruyoshi and M. Yamazaki for useful
discussions. This work is supported by the 21 COE program
``Construction of wide-angle mathematical basis focused on knots".
The work of Y.Y is supported by the Grant-in Aid for Scientific
Research (No. 17540262 and No. 17540091)
from Japan Ministry of Education. 
The work of T.O is supported by the Grant-in Aid for Scientific
Research (No. 17540262 and No. 18540285)
from Japan Ministry of Education.

\newpage


\begin{figure}[ht]
\begin{center}
\begin{minipage}{5cm}
\begin{center}
  \includegraphics[height=3cm]{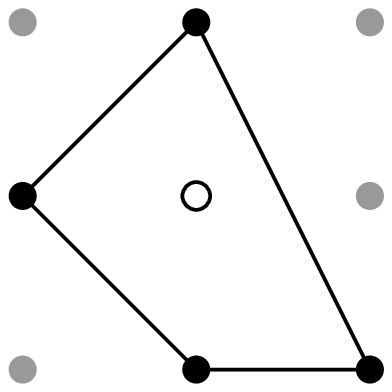} 
\end{center}
\begin{center}
  \includegraphics[height=4cm]{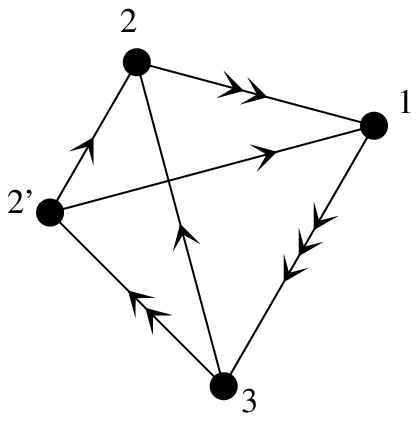}
\end{center}
\end{minipage}
\begin{minipage}{7cm}
  \begin{center}
  \includegraphics[height=8cm]{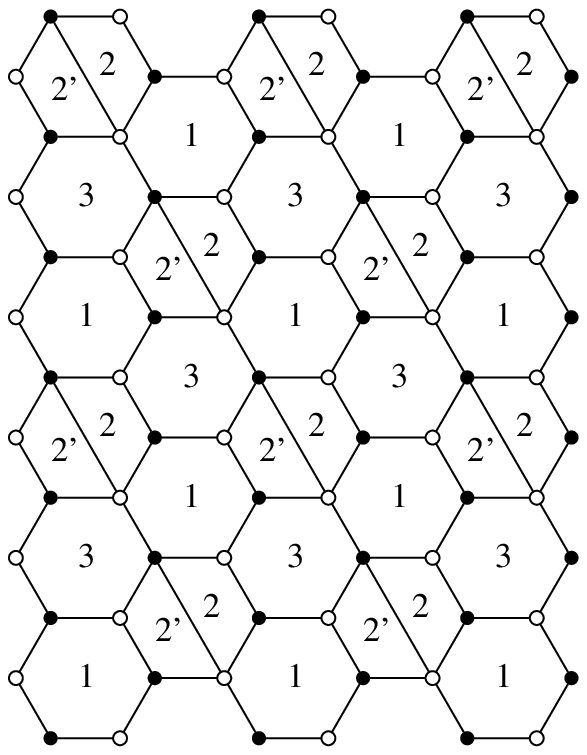} 
  \end{center}
\end{minipage}
\end{center}
\caption{Toric diagram, quiver diagram and brane tiling for $dP_1$.}
\end{figure}


\begin{figure}[ht]
\begin{center}
\begin{minipage}{5cm} 
  \begin{center}
  \includegraphics[height=3cm]{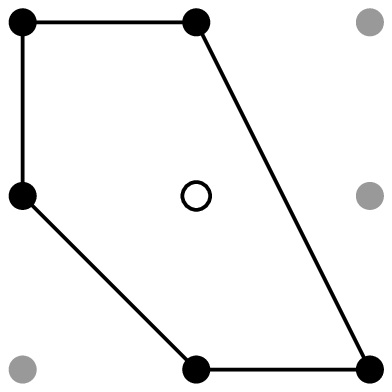} 
  \end{center}
  \begin{center}
  \includegraphics[height=4cm]{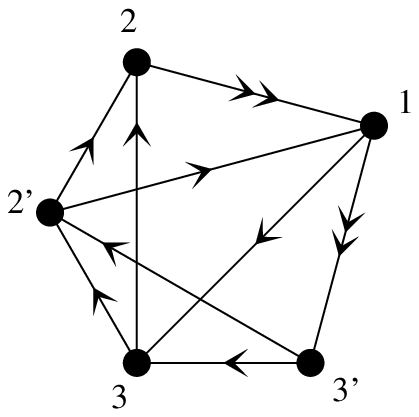}
  \end{center}
\end{minipage}
\begin{minipage}{7cm}
  \begin{center}
  \includegraphics[height=8cm]{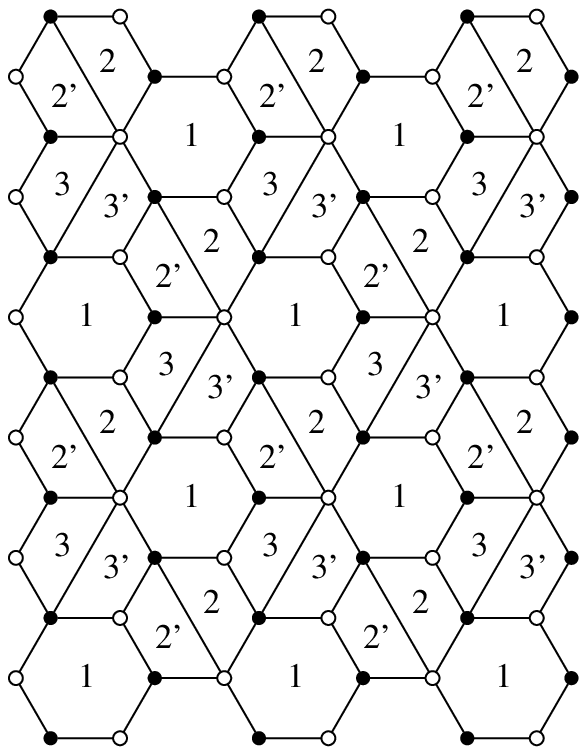} 
  \end{center}
\end{minipage}
\end{center}
\caption{Toric diagram, quiver diagram (model II) and brane tiling (model II) for $dP_2$.} 
\end{figure}
%


\begin{figure}[ht]
\begin{center} 
\begin{minipage}{5cm}
  \begin{center}
  \includegraphics[height=3cm]{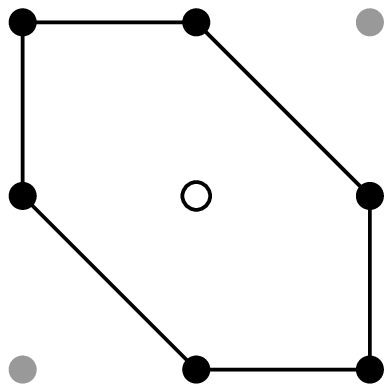} 
  \end{center}
  \begin{center}
  \includegraphics[height=4cm]{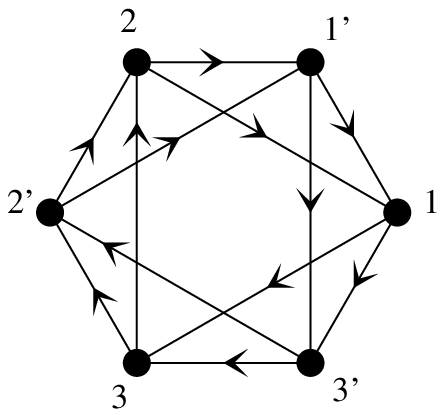}
  \end{center}
\end{minipage}
\begin{minipage}{7cm}
  \begin{center}
  \includegraphics[height=8cm]{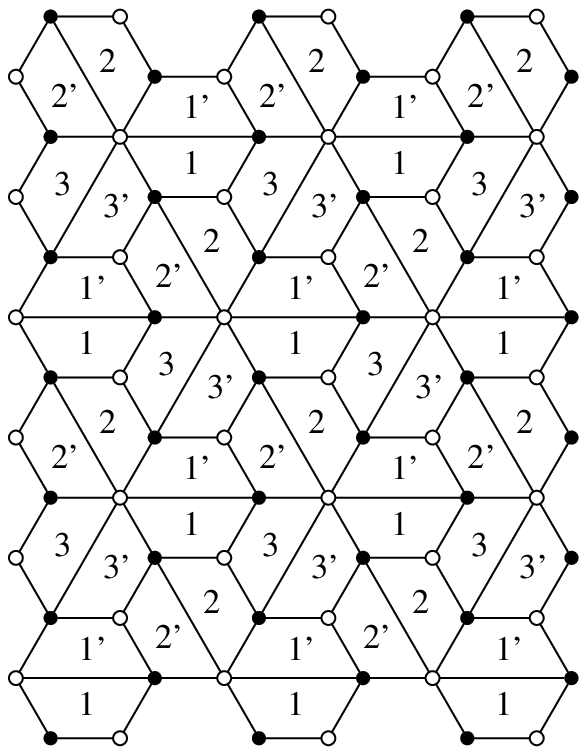} 
  \end{center}
\end{minipage}
\end{center}
\caption{Toric diagram, quiver diagram (model I) and brane tiling (model I) for $dP_3$.} 
\end{figure}%


\begin{figure}[ht]
\begin{center}
\includegraphics[height=12cm]{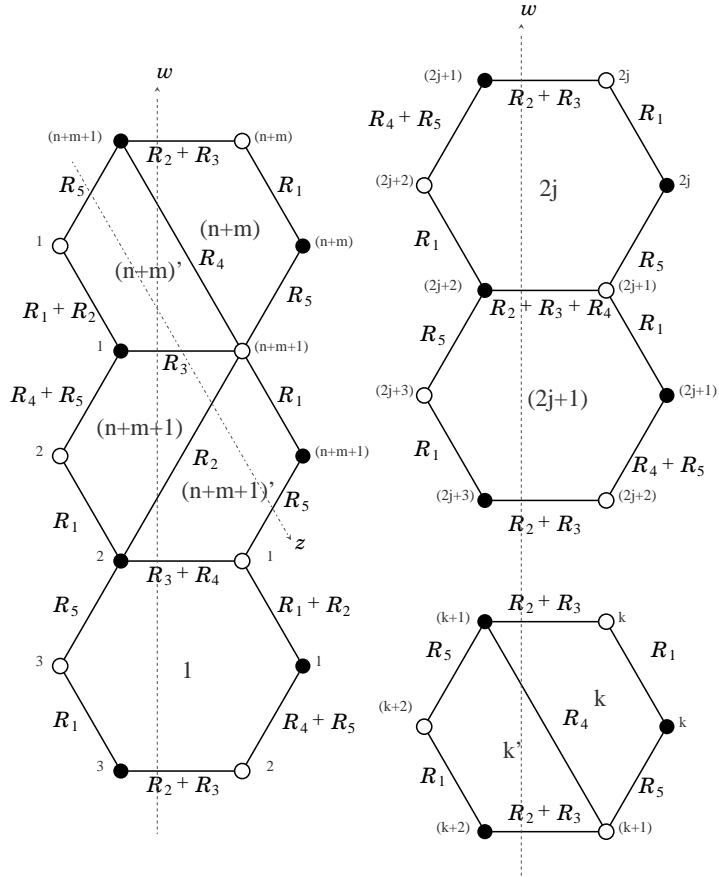}
\end{center}
\caption{Brane tiling for $X^{p,q}$.}
\label{BTileXpq}
\end{figure}


\begin{figure}[ht]
\begin{center}
\begin{minipage}{5.5cm}
  \begin{center}
  \includegraphics[height=4cm]{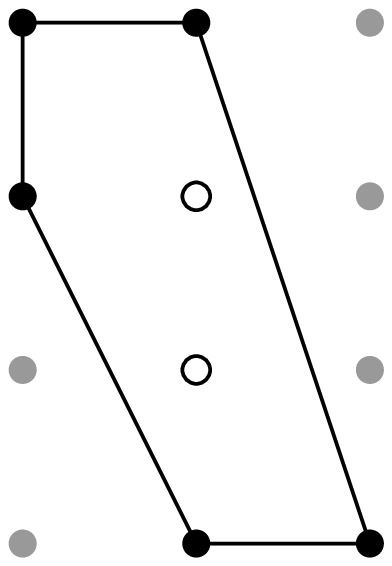} 
  \end{center}
  \begin{center}
  \includegraphics[height=5cm]{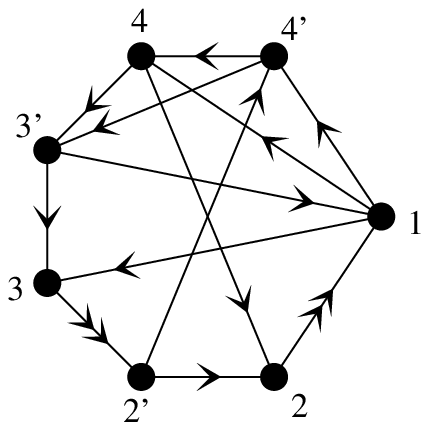}
  \end{center}
\end{minipage}
\begin{minipage}{6cm}
  \begin{center}
  \includegraphics[height=12cm]{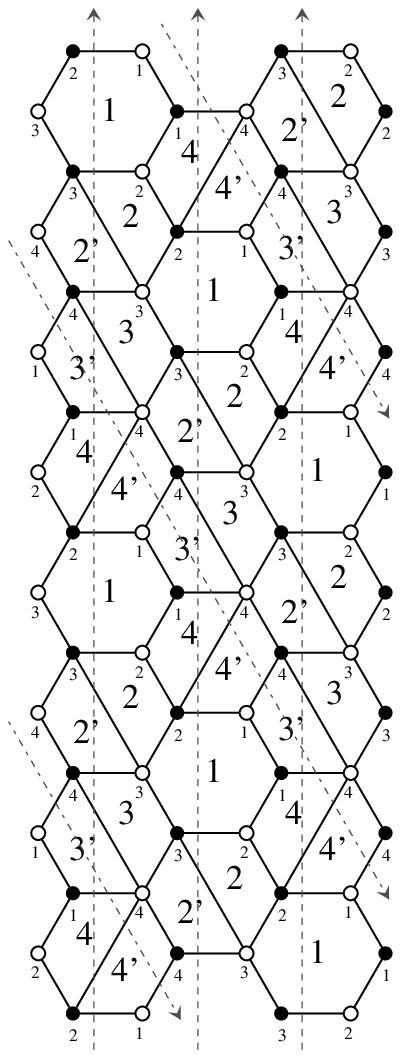} 
  \end{center}
\end{minipage}
\end{center}
\caption{Toric diagram, quiver diagram and brane tiling for $X^{3,1}$.}
\end{figure}


\begin{figure}[ht]
\begin{center}
\begin{minipage}{5.5cm}
  \begin{center}
  \includegraphics[height=4cm]{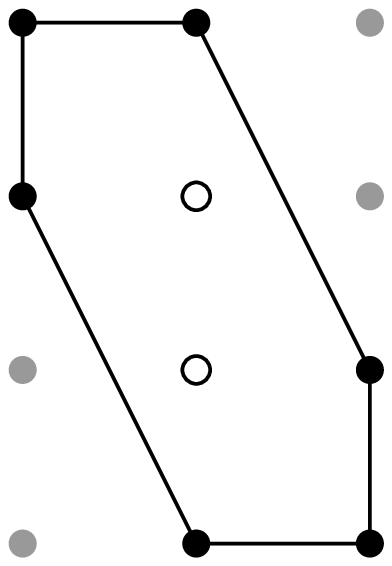} 
  \end{center}
  \begin{center}
  \includegraphics[height=5cm]{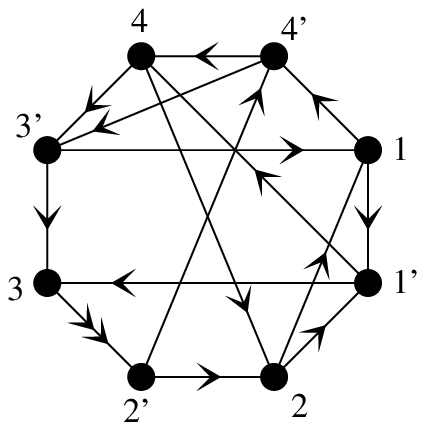}
  \end{center}
\end{minipage}
\begin{minipage}{6cm}
  \begin{center}
  \includegraphics[height=12cm]{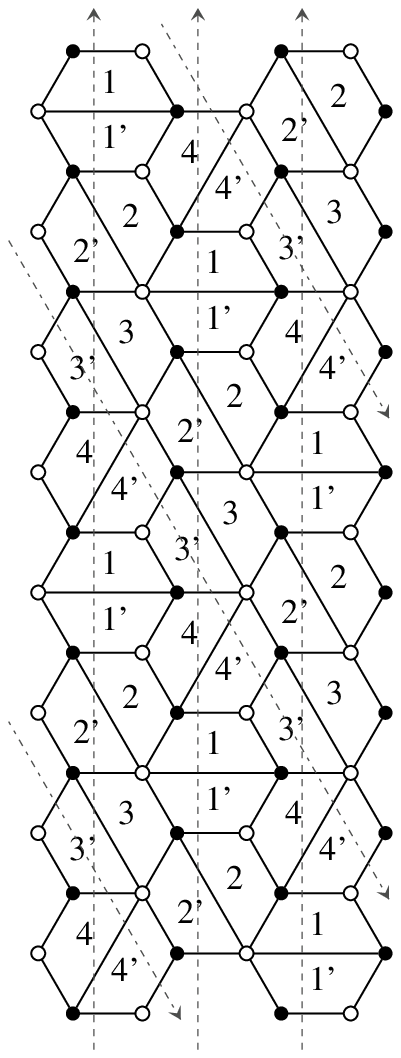} 
  \end{center}
\end{minipage}
\end{center}
\caption{Toric diagram, quiver diagram and brane tiling for $Z^{3,1}$.} 
\end{figure}


\begin{figure}[ht]
\begin{center}
\begin{minipage}{6cm}
  \begin{center}
  \includegraphics[height=15cm]{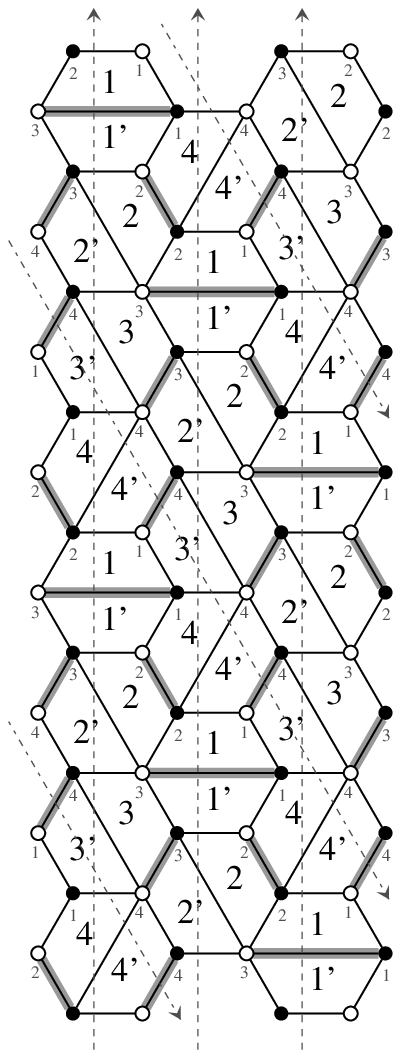}
  \end{center}
\end{minipage}
\begin{minipage}{6cm}
  \begin{center}
  \includegraphics[height=15cm]{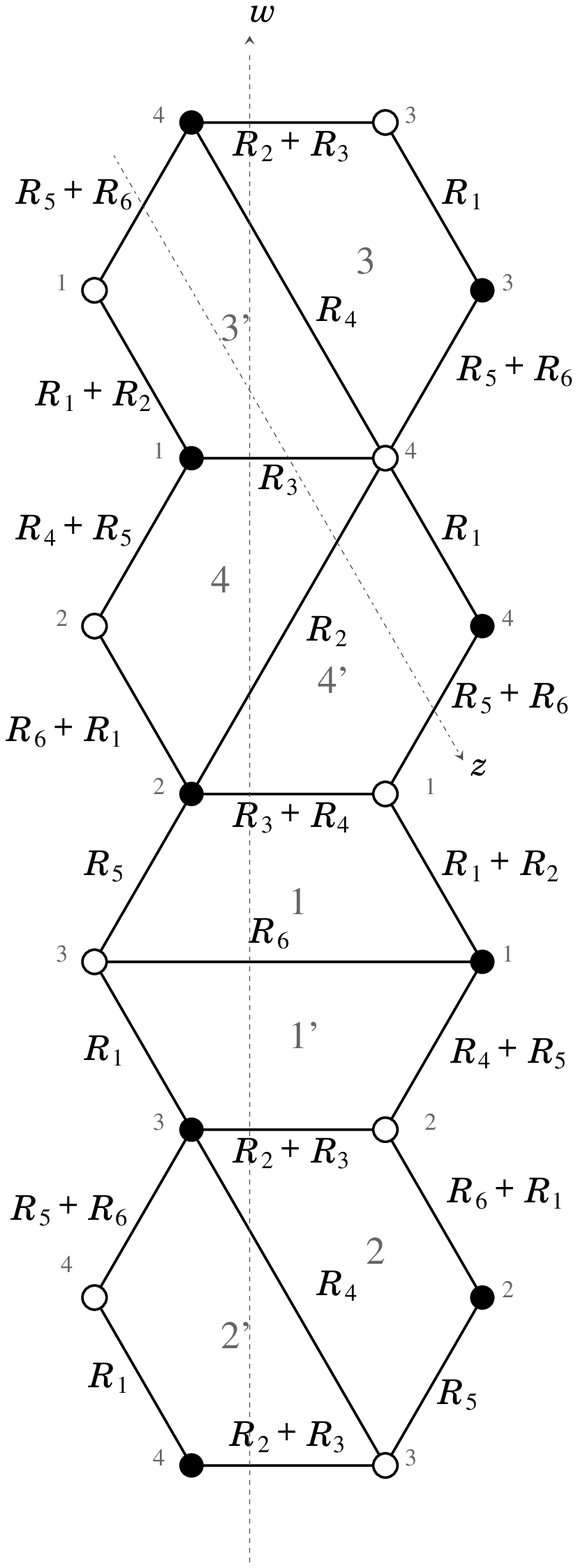} 
  \end{center}
\end{minipage}
\end{center}
\caption{The perfect matching which corresponds to the node $V_6$ 
and the $R$-charge assignment for $Z^{3,1}$.}
\end{figure}


\begin{figure}[ht]
\begin{center}
\begin{minipage}{5.5cm}
  \begin{center}
  \includegraphics[height=4cm]{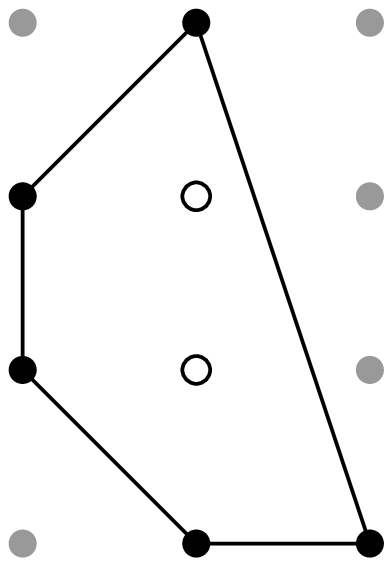} 
  \end{center}
  \begin{center}
  \includegraphics[height=5cm]{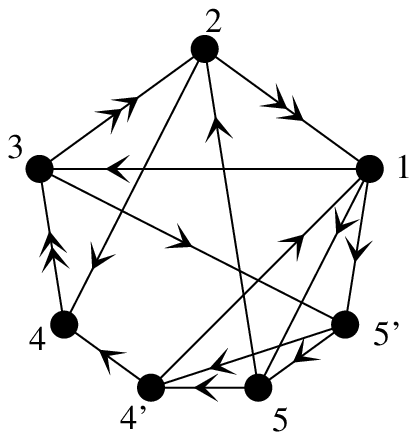}
  \end{center}
\end{minipage}
\begin{minipage}{6cm}
  \begin{center}
  \includegraphics[height=12cm]{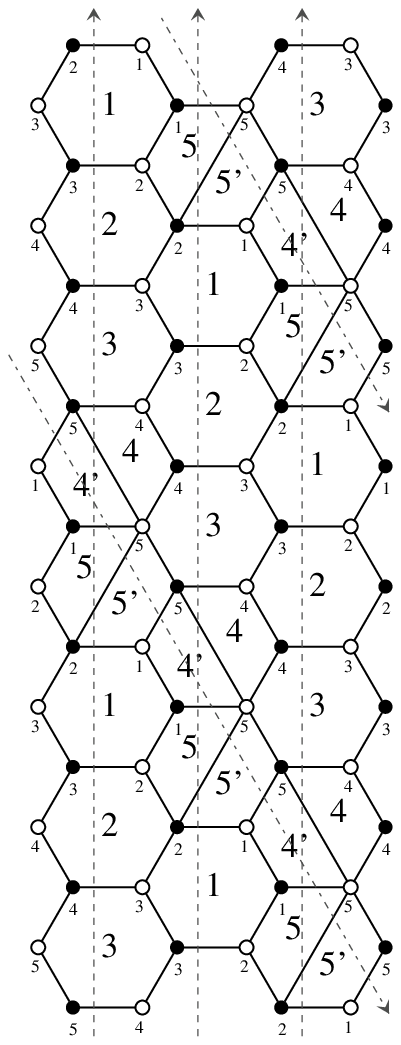} 
  \end{center}
\end{minipage}
\end{center}
\caption{Toric diagram, quiver diagram and brane tiling for $X^{3,2}$.} 
\end{figure}


\begin{figure}[ht]
\begin{center}
\begin{minipage}{5.5cm}
  \begin{center}
  \includegraphics[height=4cm]{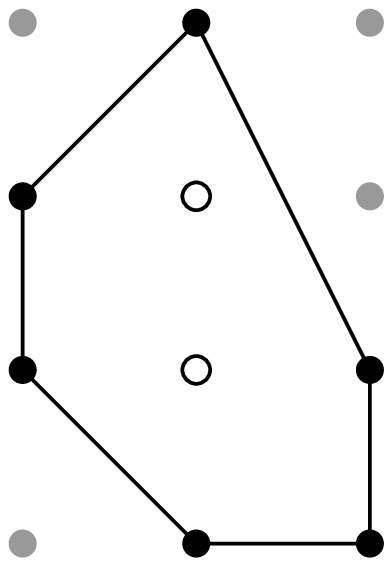} 
  \end{center}
  \begin{center}
  \includegraphics[height=5cm]{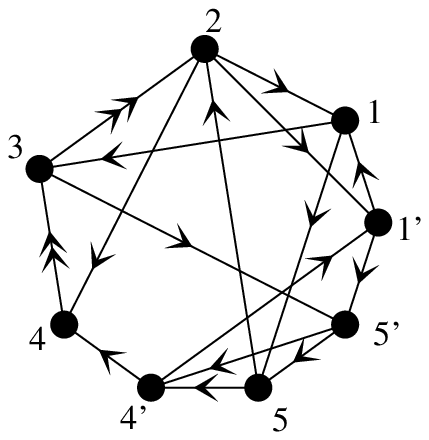}
  \end{center}
\end{minipage}
\begin{minipage}{6cm}
  \begin{center}
  \includegraphics[height=12cm]{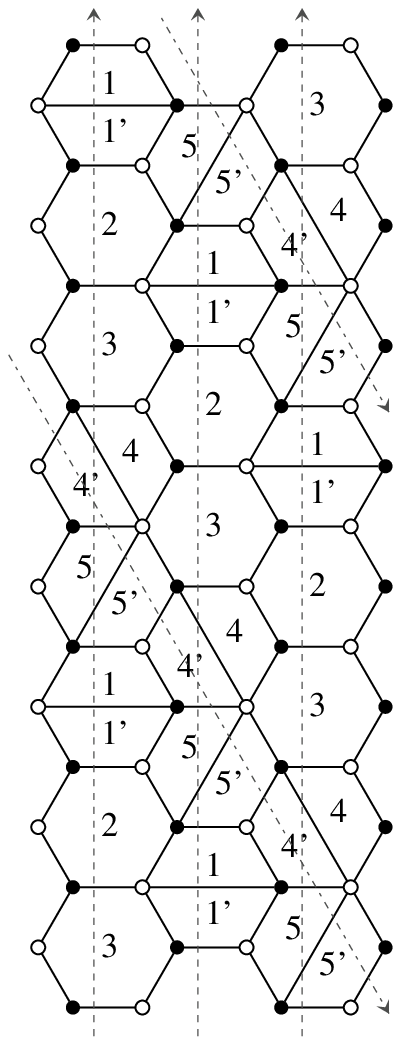} 
  \end{center}
\end{minipage}
\end{center}
\caption{Toric diagram, quiver diagram and brane tiling for $Z^{3,2}$ (case $1$).} 
\end{figure}
%


\begin{figure}[ht]
\begin{center}
\begin{minipage}{5cm}
\begin{center}
\includegraphics[height=14cm]{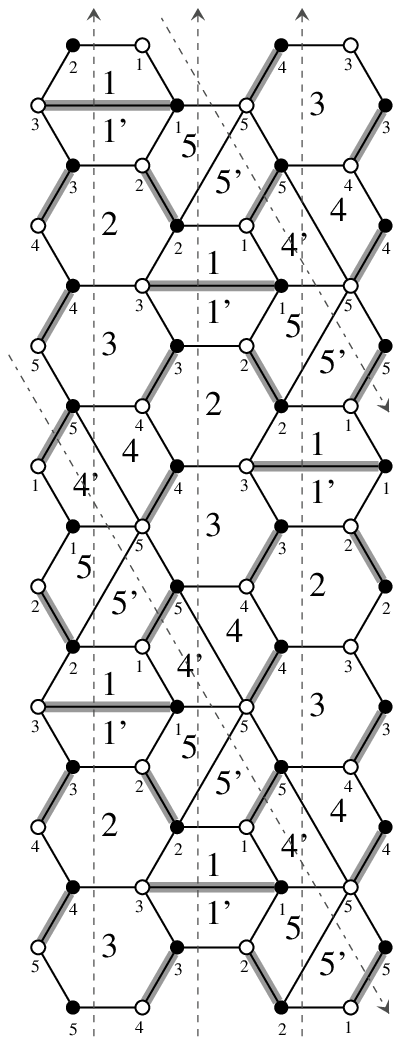} 
\end{center}
\end{minipage}
\begin{minipage}{7cm}
\begin{center}
\includegraphics[height=12cm]{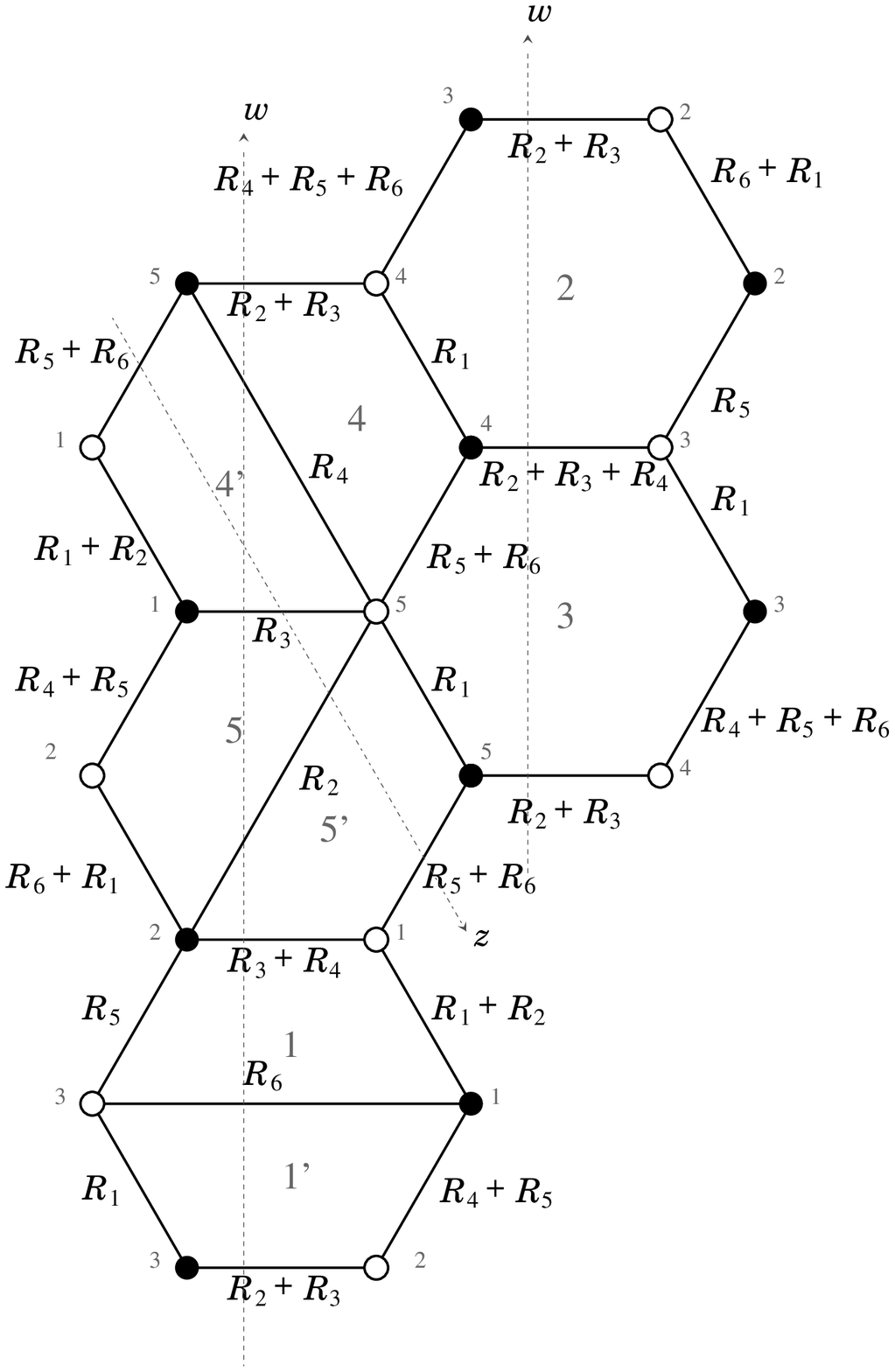} 
\end{center}
\end{minipage}
\end{center}
\caption{The perfect matching and the $R$-charge assignment for $Z^{3,2}$ (case $1$).}
\end{figure}


\begin{figure}[ht]
\begin{center}
\begin{minipage}{5.5cm}
  \begin{center}
  \includegraphics[height=4cm]{ToricZ32.eps} 
  \end{center}
  \begin{center}
  \includegraphics[height=5cm]{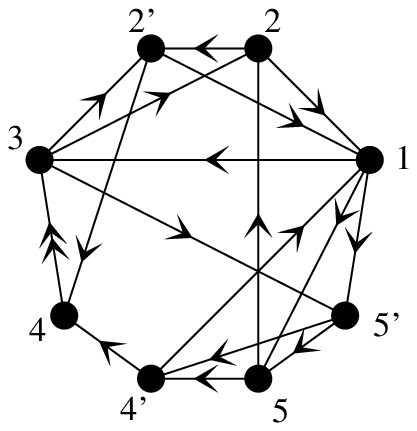}
  \end{center}
\end{minipage}
\begin{minipage}{6cm}
  \begin{center}
  \includegraphics[height=12cm]{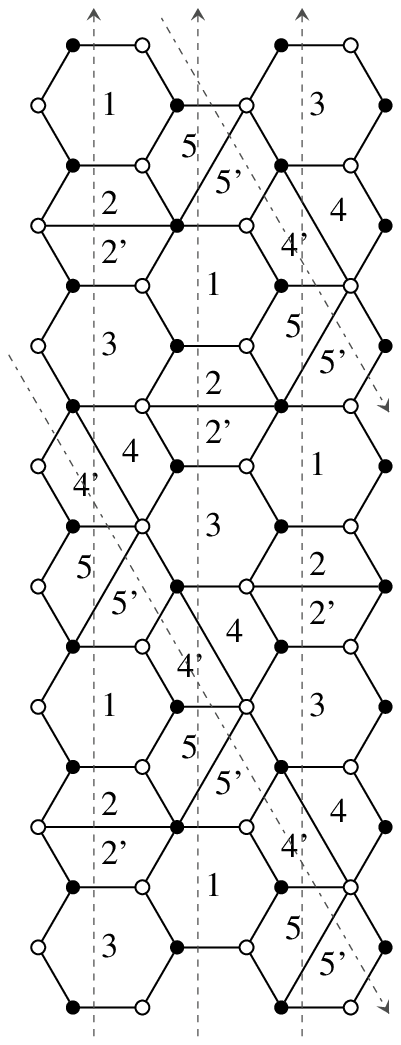} 
  \end{center}
\end{minipage}
\end{center}
\caption{Toric diagram, quiver diagram and brane tiling for $Z^{3,2}$ (case $2$).} 
\end{figure}
%


\begin{figure}[ht]
\begin{center}
\begin{minipage}{5cm}
\begin{center}
\includegraphics[height=14cm]{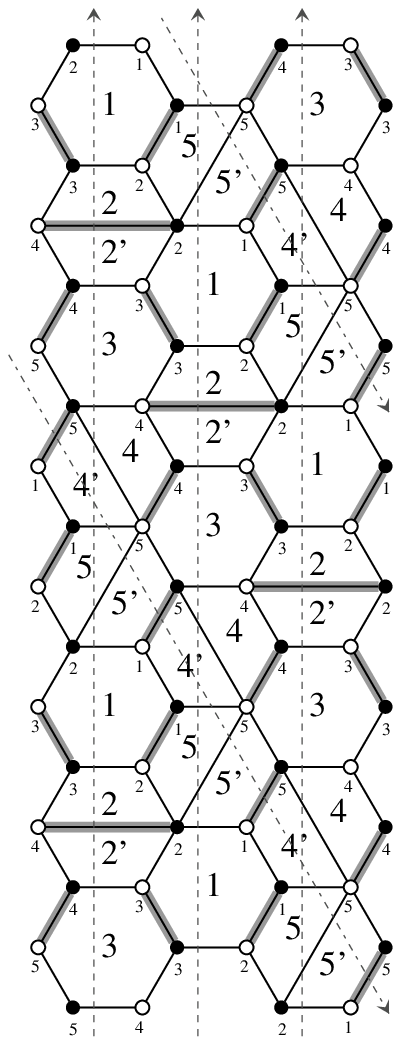} 
\end{center}
\end{minipage}
\begin{minipage}{7cm}
\begin{center}
\includegraphics[height=12cm]{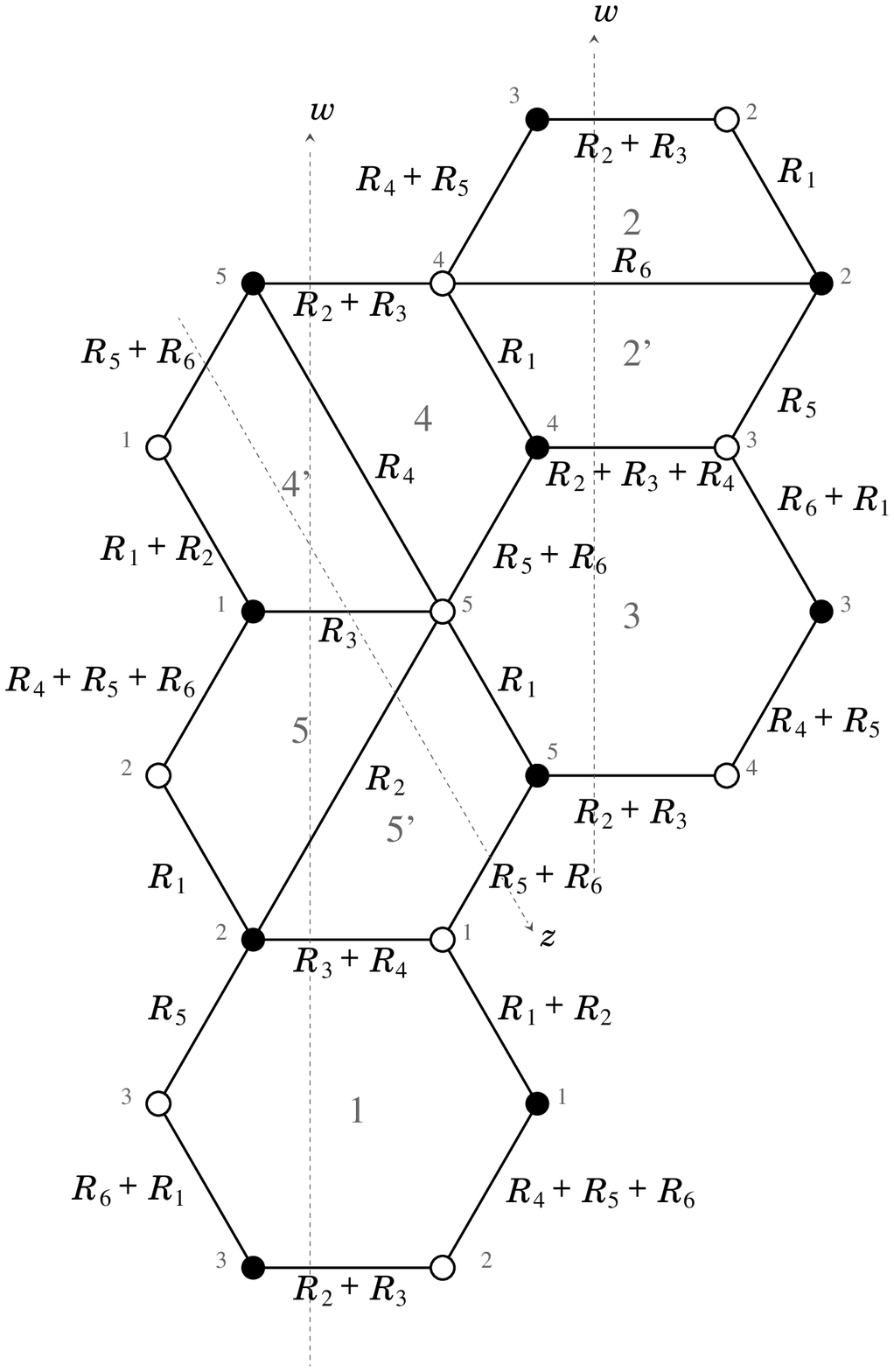} 
\end{center}
\end{minipage}
\end{center}
\caption{The perfect matching and the $R$-charge assignment for $Z^{3,2}$ (case $2$).}
\end{figure}


\begin{figure}[ht]
\begin{center}
\begin{minipage}{5.5cm}
  \begin{center}
  \includegraphics[height=4cm]{ToricZ32.eps} 
  \end{center}
  \begin{center}
  \includegraphics[height=5cm]{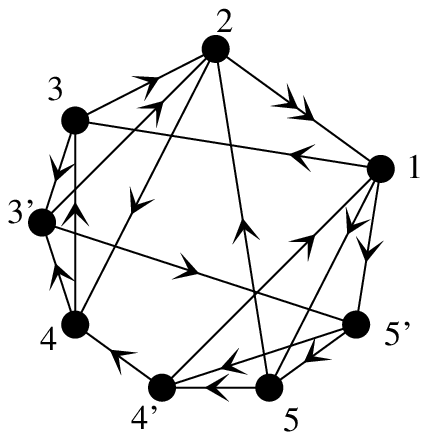}
  \end{center}
\end{minipage}
\begin{minipage}{6cm}
  \begin{center}
  \includegraphics[height=12cm]{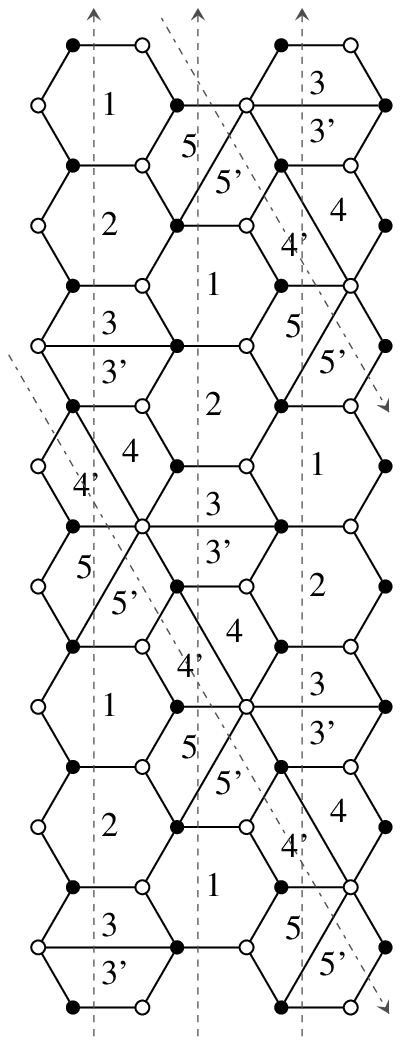} 
  \end{center}
\end{minipage}
\end{center}
\caption{Toric diagram, quiver diagram and brane tiling for $Z^{3,2}$ (case $3$).} 
\end{figure}


\begin{figure}[ht]
\begin{center}
\begin{minipage}{5cm}
\begin{center}
\includegraphics[height=14cm]{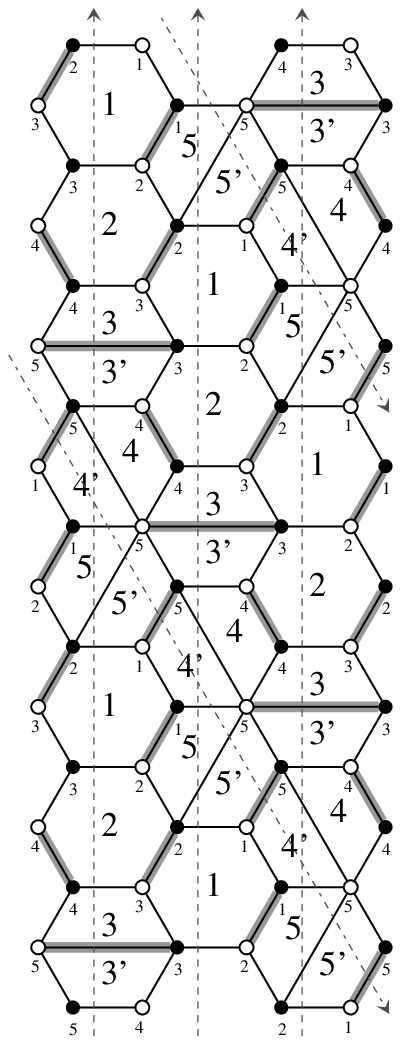} 
\end{center}
\end{minipage}
\begin{minipage}{7cm}
\begin{center}
\includegraphics[height=12cm]{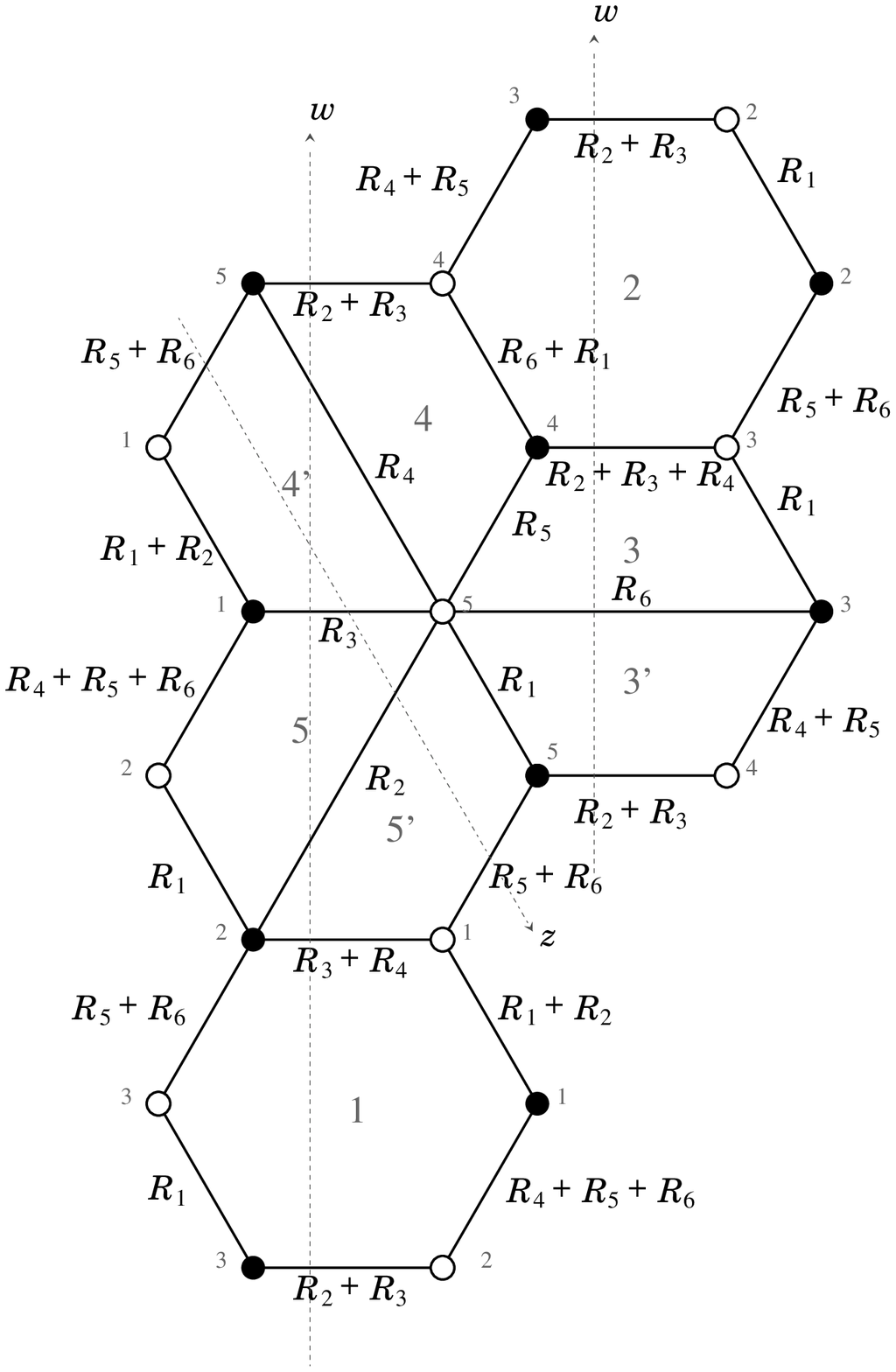} 
\end{center}
\end{minipage}
\end{center}
\caption{The perfect matching and the $R$-charge assignment for $Z^{3,2}$ (case $3$).}
\end{figure}


\begin{figure}[ht]
\label{TDXandZ}
\begin{center}
\begin{minipage}{7cm}
  \begin{center}
  \includegraphics[height=6cm]{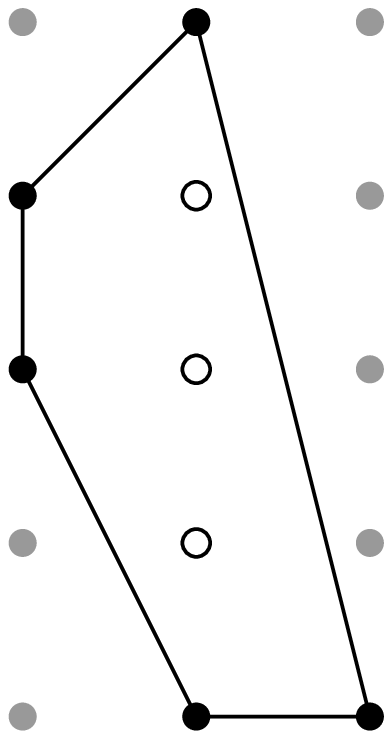} 
  \end{center}
  \caption{Toric diagram for $X^{p,q}$.}
\end{minipage}
\begin{minipage}{7cm}
  \begin{center}
  \includegraphics[height=6cm]{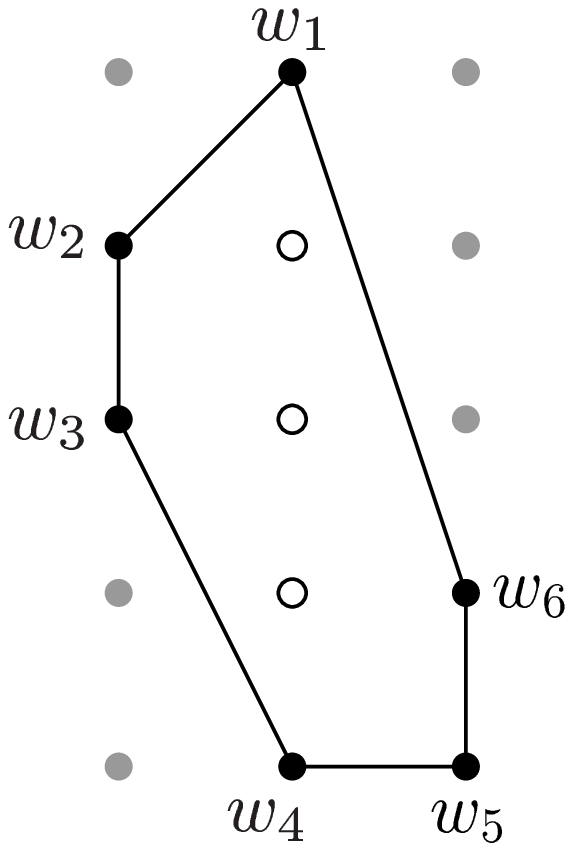} 
  \end{center}
  \caption{Toric diagram for $Z^{p,q}$.}
\end{minipage} 
\end{center}
\end{figure}%

\end{document}